# Strong in-plane magnetic field induced reemergent superconductivity in the van der Waals heterointerface of NbSe$_2$ and CrCl$_3$

*Da Jiang[†,#,∥,\*], Tianzhong Yuan[‡,∥], Yongzheng Wu[‡,∥], Xinyuan Wei[‡], Gang Mu[†], Zhenghua An[‡,§],*

*Wei Li[‡,§,\*]*

[†]State Key Laboratory of Functional Materials for Informatics, Shanghai Institute of

Microsystem and Information Technology, and Center for Excellence in Superconducting

Electronics, Chinese Academy of Science, Shanghai 200050, Peoples Republic of China.

[‡]State Key Laboratory of Surface Physics and Department of Physics, Fudan University,

Shanghai 200433, Peoples Republic of China.

[§]Collaborative Innovation Center of Advanced Microstructures, Nanjing 210093, China.

[#]Center of Materials Science and Optoelectronics Engineering, University of Chinese Academy

of Sciences, Beijing 100049, China.

ABSTRACT

The magnetic field is generally considered to be incompatible with superconductivity as it tends to spin-polarize electrons and breaks apart the opposite-spin singlet superconducting Cooper pairs. Here an experimental phenomenon is observed that an intriguing reemergent superconductivity evolves from a conventional superconductivity undergoing a hump-like intermediate phase with a finite electric resistance in the van der Waals heterointerface of layered NbSe$_2$ and CrCl$_3$ flakes. This phenomenon merely occurred when the applied magnetic field is parallel to the sample plane and perpendicular to the electric current direction as compared to the reference sample of NbSe$_2$ thin flake. The strong anisotropy of reemergent superconducting phase is pointed to the nature of Fulde-Ferrell-Larkin-Ovchinnikov (FFLO) state driven by the strong interfacial spin-orbit coupling between NbSe$_2$ and CrCl$_3$ layers. The theoretical picture of FFLO state nodes induced by Josephson vortices collectively pinning is presented for well understanding the experimental observation of the reemergent superconductivity. This finding sheds light on an opportunity to search for the exotic FFLO state in the van der Waals heterostructures with strong interfacial spin-orbit coupling.

INTRODUCTION

The study of unconventional superconductivity (SC) with a nontrivial Cooper pairing order parameter has been one of the most challenging subjects in both condensed matter physics and material science[1-3]. Usually, unconventional SC is characterized by anisotropic superconducting gap functions, which change sign in momentum space, such as $d$-wave and $s^{\pm}$-wave pairing order parameters. These pairing order parameters have been observed in the classes of high-temperature ($T_c$) superconductors of cuprates[4-6] and iron-based superconductors[7-10].

A novel SC that belongs to another class of unconventional superconducting state was first theoretically predicted by Fulde and Ferrell (FF)[11], and Larkin and Ovchinnikov (LO)[12] in a conventional *s*-wave superconductor in the presence of Zeeman field over fifty years ago. In the FFLO state[13-17], the superconducting order parameter changes sign in real space. Such exotic superconducting state originates from the paramagnetism of conduction electrons, giving rise to the Cooper pair formation occurring between Zeeman split parts of the Fermi surface and a new superconducting pairing state with finite momentum. As a result, the Cooper pairs have finite center of mass momentum, leading to an oscillating part of the superconducting order parameter in real space with wavelength in the order of coherence length. One typical candidate for the emergence of the FFLO state is the heavy-fermion compound CeCoIn$_5$[13-18]. Subsequently, the neutron-diffraction experiments showed that the superconducting and magnetic ordering phenomena are coupled with each other[19-20], making the nature of such heavy fermion system to be elusive. Good evidence for an existence of the FFLO state has been demonstrated in organic superconductors using Nuclear Magnetic Resonance[21-23] and heat capacity[24-26] with fairly stringent condition of applied external magnetic field up to 45 tesla[27].

Here we propose that the exotic FFLO state could be realized in the NbSe$_2$/CrCl$_3$ van der Waals heterostructure based on the nanofabrication and the electric transport measurements [Figure 1(a)]. The layered antiferromagnetic insulating material, chromium trihalides (CrCl$_3$)[28,29], and the layered type-II superconductor 2H-NbSe$_2$[30-32] are easy to form a magnetic-superconductor heterostructure[33] with the close association of atomic arrangement [Figure 1(b)]. Its interfacial interaction induced strong spin-orbit coupling will drive the system into the possible FFLO state in NbSe$_2$ in the presence of in-plane magnetic field[34]. Firstly, the reemergent SC is merely experimentally observed in the NbSe$_2$/CrCl$_3$ heterostructure well above the field of 4 tesla as compared to the reference sample of NbSe$_2$ thin flake, which evolves from a conventional SC undergoing a hump-like intermediate phase with a finite electric resistance when the magnetic field is applied parallel to the sample plane and perpendicular to the electric current direction. Secondly, the hump-like phase will disappear and display a SC when the magnetic field orientation is changed along the electric current direction. Thirdly, the reemergent SC is invisible for the out-of-plane magnetic field orientations. These strong anisotropic behaviors rule out the possibilities from the mechanisms of disorder induced flux-pinning magnetoresistance (MR)[35,36] and exchange field compensation effect[37,38], implying the nature of FFLO state for the reemergent superconducting phase driven by the strong interfacial spin-orbit coupling. A theoretical picture of FFLO nodes induced by collectively flux-pinning[27] is then presented for well understanding the experimental observation of the reemergent SC.

EXPERIMENTAL RESULTS

The NbSe$_2$/CrCl$_3$ van der Waals heterostructure [Figure 1(a) and Figure S2 in Supporting Information (SI)] is fabricated to study the FFLO state in its interface by electric transport measurements (see SI for the experimental details). Atomic force microscope measurements show that the layered step height is 90 nm and 340 nm for the NbSe$_2$ and CrCl$_3$ flakes, respectively. A reference sample of NbSe$_2$ thin flake exhibits a sharp superconducting transition with a critical temperature $T_c \approx 6.83$ K [Figure 1(c)], consistent with previous studies[32,39]. The $T_c$ is defined as the temperature at the midpoint of the transition in electric resistance. Comparing with the NbSe$_2$ reference sample, the superconducting critical temperature of the NbSe$_2$/CrCl$_3$ heterostructure almost remains the same value of 6.92 K [Figure 1(c)]. This result is reasonable since the CrCl$_3$ is an antiferromagnetic layered insulator[28], indicating the absence of charge transfer between NbSe$_2$ and CrCl$_3$. Such indication is also confirmed by the first-principles calculations that the calculated density of state for CrCl$_3$ displays an insulator with an energy gap of 1.66 eV (see Figure S11 in SI).

The MR measurements are then carried out on both the NbSe$_2$ reference sample and the NbSe$_2$/CrCl$_3$ heterostructure in the presence of vertical magnetic field orientation shown in Figure 1(d) & 1(e), respectively. The fundamental superconducting behaviors are observed for both the NbSe$_2$ reference sample and the NbSe$_2$/CrCl$_3$ heterostructure. The superconducting critical magnetic field ($H_{c2}^{mid}$) shifts to a lower value with increasing temperature, where $H_{c2}^{mid}$ is defined as the magnetic field at the midpoint of the transition in electric resistance. The mechanisim[40] is that the orbital effect induced by the vertical magnetic field leads to the appearance of Abrikosov vortices and forms a regular array of vortex lines parallel to the magnetic field. As a result, the kinetic energy of superconducting currents around the vortex cores reduces the superconducting condensation energy[40]. Figure 1(f) shows a quantitative estimation of $H_{c2}^{mid}$ for the out-of-plane orientation as a function of temperature. The solid fitting lines are the expected dependence for a superconductor with a linear formula of two-dimensional (2D) Ginzburg-Landau model[41-45] $H_{c2,\perp}^{mid} = \frac{\Phi_0}{2\pi\xi^2}\left(1 - \frac{T}{T_c}\right)$, where $\Phi_0$ is the flux quantum and $\xi$ is the superconducting coherence length. From the extrapolation of $H_{c2,\perp}^{mid}$ at $T$ = 0 K, we extract the orbital limiting pair breaking field $H_{c2}^{orb} = \frac{\Phi_0}{2\pi\xi^2} = 4.6$ T for the NbSe$_2$ reference sample and $H_{c2}^{orb} = 5.29$ T for the NbSe$_2$/CrCl$_3$ heterostructure, which is much larger than that for the NbSe$_2$ reference sample[44]. The enhancement of orbital limiting field mainly stems from the significant contribution of Rashba-type spin-orbit coupling induced by the interfacial electric field along the out-of-plane orientation[42-45], resulting in increasing the anisotropic superconducting energy gap[45] and the reduction of in-plane superconducting coherence length $\xi$. The estimation of in-plane superconducting coherence length are $\xi \approx 8.46$ nm and 7.89 nm for the NbSe$_2$ reference sample and the NbSe$_2$/CrCl$_3$ heterostructure, respectively. The MR measurements for the in-plane magnetic field orientations are presented in Figure S4C & D in SI.

The out-of-plane polar angle ($\theta$) dependence of the critical magnetic field $H_{c2}^{mid,\theta}$ at a fixed temperature of 3 K is systemically investigated to verify the 2D behavior of the superconducting layer. Figure 2(a) shows the MR measurements at various polar angle values $\theta$ ranging from $\theta = 0°$ to $90°$. The critical magnetic field $H_{c2}^{mid,\theta}$ as a function of polar angle $\theta$ is extracted in Figure 2(b). Furthermore, the data are fitted with the 2D Tinkham formula[46] (red solid curve) and the three-dimensional (3D) anisotropic Ginzburg-Landau (G-L) model[42] (green solid curve), given by $\frac{H_{c2}^{mid,\theta}|\cos\theta|}{H_{c2,\perp}^{mid}} + \left(\frac{H_{c2}^{mid,\theta}\sin\theta}{H_{c2,\parallel}^{mid}}\right)^2 = 1$ and $\left(\frac{H_{c2}^{mid,\theta}\cos\theta}{H_{c2,\perp}^{mid}}\right)^2 + \left(\frac{H_{c2}^{mid,\theta}\sin\theta}{H_{c2,\parallel}^{mid}}\right)^2 = 1$, respectively. For an overall range, the data seems to be well described by both theoretical models. However, a closer look at the region around $\theta = 90°$ in Figure 2(b), and considering the strong anisotropy of MR measurements for the NbSe$_2$/CrCl$_3$ heterostructure and the fact that the angular dependence of $H_{c2}^{mid,\theta}$ for NbSe$_2$ has been found to follow the 2D Tinkham model[47], the tendency of experimental data prefers to follow the 2D Tinkham model rather than the 3D Ginzburg-Landau model. In addition, it is interesting to point out that the anomalous MR curves appear at $\theta = 85°$ and $80°$, where a "dip structure" appears at a high magnetic field in the NbSe$_2$/CrCl$_3$ heterostructure, i.e. the electric resistance decreases as increasing the magnetic field at around 8

T at $\theta = 85°$. To clarify the behavior of "dip structure" in the vicinity of in-plane magnetic field in the NbSe$_2$/CrCl$_3$ heterostructure, the sample is rotated to 90° making the magnetic field direction parallel to the sample plane but perpendicular to the electric current direction. For a global view on the electric resistance as a function of temperature at various strengths of magnetic field, the $T_c$ shifts to a lower temperature value as increasing the magnetic field strength [Figure 3(a)]. The fascinating "dip structure" is visible at the strength of magnetic field ranging from H = 3 T to 7 T. More interestingly, when the strength of magnetic field is fixed at H = 7 T, the electric resistance not only displays a "dip structure", but also reenters into a new superconducting state at the low temperature. Furthermore, such an anomalous phenomenon also exists in the MR measurements [Figure 3(b)], i.e. at a fixed temperature of 3.4 K, the electric resistance firstly displays a conventional superconducting behavior with zero value of electric resistance in the absence of external magnetic field. When the strength of in-plane magnetic field is gradually increased, the superconducting state will be destroyed and the system will enter into a hump-like phase displaying a finite value of electric resistance. As further increasing the strength of in-plane magnetic field, however, the electric resistance drops down and reenters into a second superconducting state. When the strength of magnetic field exceeds $H_{c2,\parallel}^{mid}$, the second SC will be destroyed completely, and then the system eventually enters into a conventional Fermi liquid state, consistent with the previous studies on the NbSe$_2$[44]. On the other hand, the electric transport measurements of MR and R-$T$ at various in-plane azimuthal angle $\varphi$ on the NbSe$_2$ reference sample does not display any "dip structure" nor reemergent superconducting phase (Figure S3 to S5 in SI), implying that the reemergent superconducting state in the NbSe$_2$/CrCl$_3$ heterostructure is tightly related to the strong interfacial coupling between the CrCl$_3$ and NbSe$_2$ layers, such as the strong interfacial spin-orbit coupling in the NbSe$_2$/CrCl$_3$ heterostructure[34].

To further clarify the behaviors of reemergent superconducting state in the presence of strong in-plane magnetic field, the in-plane azimuthal angle $\varphi$ dependent MR measurements are performed and shown in Figure 3(c), which displays a strong anisotropy with an almost two-fold rotational symmetry. Figure 3(d) shows the detailed phase diagram of $H_{c2}^{zero} - T$ with the magnetic field parallel to the sample plane but perpendicular to the electric current direction ($\theta$=90°, $\varphi$=90°), where $H_{c2}^{zero}$ is defined at the points of zero value of electric resistance as a function of temperature $T$ in MR measurements. There are three phases in the phase diagram of $H_{c2}^{zero} - T$, including a conventional superconductivity at a weak strength of magnetic field (the pink area), the strong anisotropic hump-like intermediate phase (the blue area), and a mystery phase, namely the reemergent SC, at a strong strength of in-plane magnetic field and a low temperature (the yellow area).

DISCUSSIONS

Theoretically, to the best of our knowledge, the nature of intriguing reemergent superconducting phase observed in the NbSe$_2$/CrCl$_3$ heterostructure in the in-plane magnetic field could be understood through the mechanisms of that (i) the disorder induced flux-pinning nature[35,36]; (ii) the exchange field compensation effect[37,38]; (iii) the triplet pairing SC induced by the enhancement of interfacial spin-orbit coupling[48,49]; and (iv) the FFLO state[11,12]. Since the hump-like phase and reemergent superconducting region change dramatically by switching the electric current flowing direction from positive to negative with fixing the magnetic field parallel to the sample plane and perpendicular to the electric current direction [the azimuthal angle changes from $\varphi=90°$ to $270°$ while $\theta$ keeps $90°$ as shown in Figure 3(e)], it easily rules out the possibility of the disorder induced flux-pinning mechanism. The second possibility of the exchange field compensation mechanism could also be ruled out because the observed reemergent superconducting phase is evolved from a conventional superconducting state by tuning the in-plane magnetic field without undergoing an exchange field compensating to a negative internal exchange field. Meanwhile, the appearance of hump-like phase locates below the critical magnetic field $H_{c2}$ when the magnetic field is tuned perpendicular to the electric current direction. Considering the reemergent superconducting phase adiabatically evolved from a conventional superconducting state without undergoing a quantum phase transition, it leads the triplet Cooper pairing to be unfavorable[48,49]. These findings are quite different to the similar phenomena of hump-like phase observed in heavy fermion superconductors with time-reversal symmetry breaking triplet pairing[50-52] and in the superconducting $Mo_{0.79}Ge_{0.21}$ nanostrip intertwined with vortex and defects[53-55]. Therefore, the experimental observation of anisotropic reemergent superconducting phase is pointed to the nature of FFLO state, which is driven by the strong interfacial spin-orbit coupling in the $NbSe_2/CrCl_3$ heterostructure with close atomic arrangement in the presence of in-plane magnetic field[34].

In addition, the phenomenological theoretical model with a tunable Rashba-type spin-orbit coupling induced by the interfacial electric field between the $NbSe_2$ and $CrCl_3$ layers is also carried out for further clarifying the experimental observations in the phase diagram of $H_{c2}^{zero} - T$ in Figure 3(d) in the $NbSe_2/CrCl_3$ heterostructure (see details in SI). The calculated theoretical results suggest that the appearance of FFLO state with periodic modulation of pairing order parameters in real space could be easily driven by the spin-orbit coupling in the presence of in-plane magnetic field [see Figure 3(f) and Figure S12 in SI], consistent with previous theoretical proposal in LaAlO$_3$/SrTiO$_3$[34]. This calculation further supports the FFLO state nature for the reemergent SC in the NbSe$_2$/CrCl$_3$ heterostructure.

At last, we present a general picture to understand the experimental observations of intriguing reemergent superconducting phase in the van der Waals heterostructure of NbSe$_2$/CrCl$_3$ in the presence of in-plane magnetic field. When the magnetic field is applied in the in-plane orientation but perpendicular to the electric current direction, at low fields, the electric current flows through the Josephson vortices[27,56-58] induced by the applied in-plane magnetic field and drives the Josephson vortices flowing, developing a finite value of electric resistance. In the case of a magnetic field parallel to the sample plane along the electric current direction, however, there is absent of Josephson vortices flowing, remaining the zero value of electric resistance[57,58]. This is the nature of the hump-like intermediate phase locating inside the critical magnetic field $H_{c2}$ in Figure 3(d). If the magnetic field is further increased, the system enters into a FFLO state and the Josephson vortices collectively pin inside the nodes of FFLO state, resulting in zero value of electric resistance in the electric transport measurement. Thus, the intriguing FFLO state could be hosted in the NbSe$_2$/CrCl$_3$ van der Waals heterostructure with strong interfacial spin-orbit coupling by tuning the applied in-plane magnetic field.

CONCLUSION

We carry out the electrical transport measurmeents on the van der Waals heterostructure of NbSe$_2$/CrCl$_3$ synthesized by the nanofabrication, and find an intriguing anisotropic reemergent superconducting phase when the applied magnetic field is parallel to the sample plane, implying the nature of FFLO pairing state for observing the reemergent superconducting phase, which is driven by the strong interfacial spin-orbit coupling between NbSe$_2$ and CrCl$_3$ layers in the presence of strong in-plane magnetic field. This finding not only sheds light on the studies of unconventional superconductivity, but also opens an avenue to exploring novel quantum states in the van der Waals heterostructures with strong interfacial spin-orbit coupling based on the nanofabrication.

FIGURES

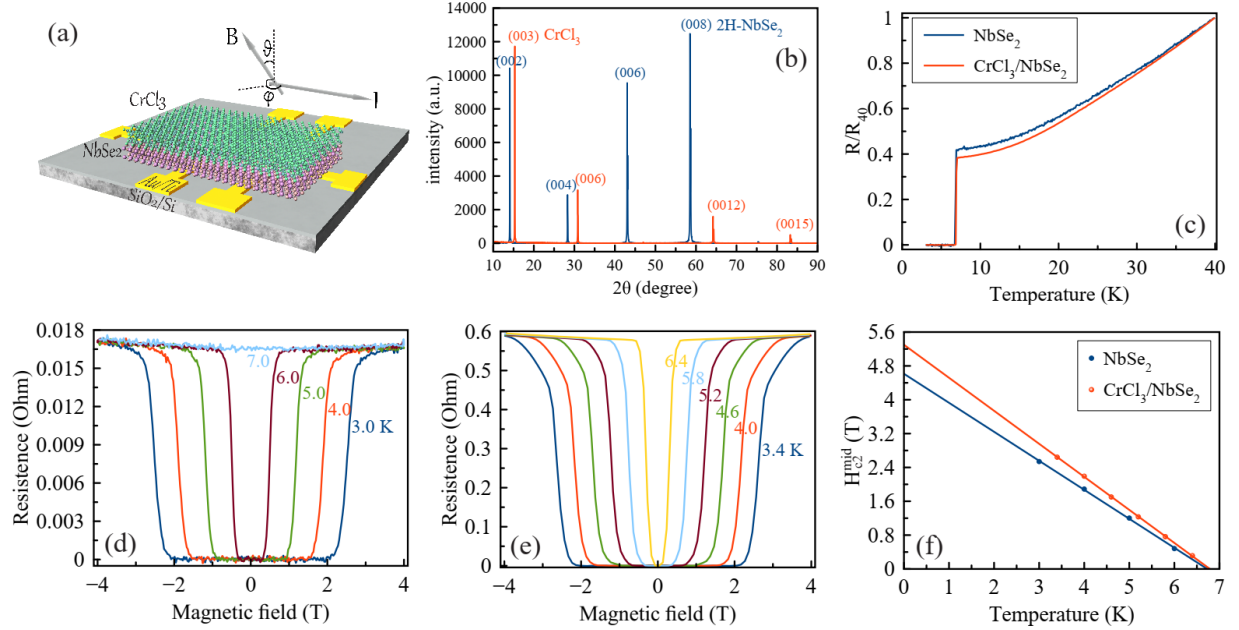

**Figure 1.** (a) The schematic illustration of the NbSe$_2$/CrCl$_3$ hterostructure. The magnetic field shows in spherical coordinate with polar angle $\theta$ and azimuthal angle $\varphi$. The *x*- and *y*-axes are defined in the in-plane orientations ($\theta$ =90°) with $\varphi$= 0° and 90°, respectively. The *z*-axis is defined as the out-of-plane orientation ($\theta = 0°$). (b) The XRD pattern of single crystal of NbSe$_2$ with 2H type structure and the CrCl$_3$. (c) The rescaled electric resistance as a function of temperature curves for the NbSe$_2$/CrCl$_3$ hterostructure and the NbSe$_2$ reference sample. The R$_{40}$ denotes the electric resistance measured at the temperature of 40 K. (d) and (e) The MR curves of the NbSe$_2$ reference sample and the NbSe$_2$/CrCl$_3$ hterostructure with various strengths of vertical magnetic field ($\theta = 0°$). (f) $H_{c2}^{mid}$ values extracted from (d) and (e) as a function of temperature. The solid lines are linear fitted to $H_{c2}^{mid}$.

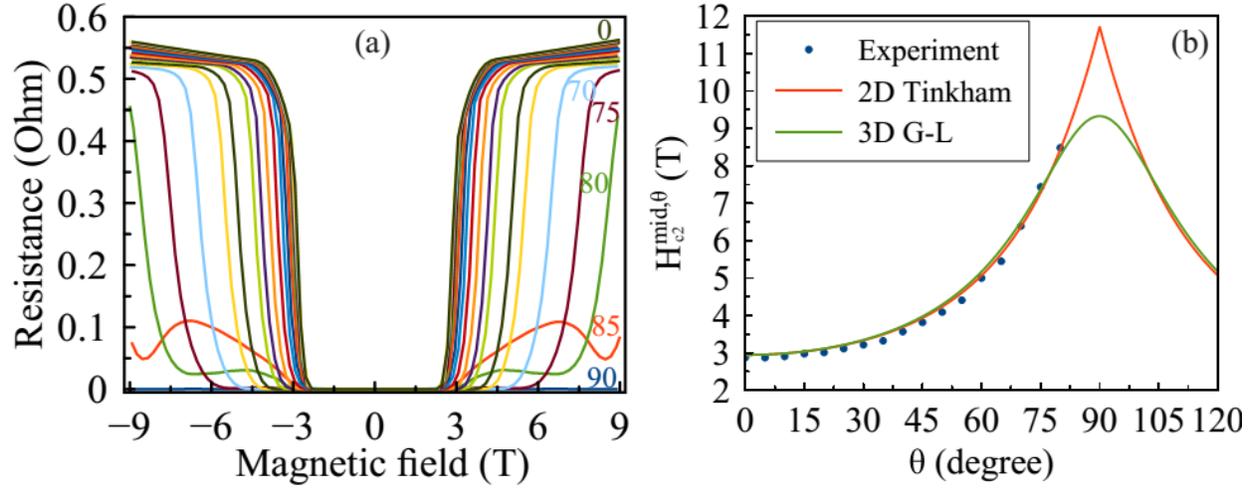

**Figure 2.** (a) The electric resistance as a function of the strength of magnetic field at various polar angle values $\theta$ ranging from $\theta = 0°$ to $90°$ at fixed temperature of 3 K and $\varphi = 0°$. (b) The critical magnetic field $H_{c2}^{mid,\theta}$ as a function of polar angle $\theta$. The red solid and green solid lines are fitting to the data by using the 2D Tinkham model and 3D G-L model model, respectively.

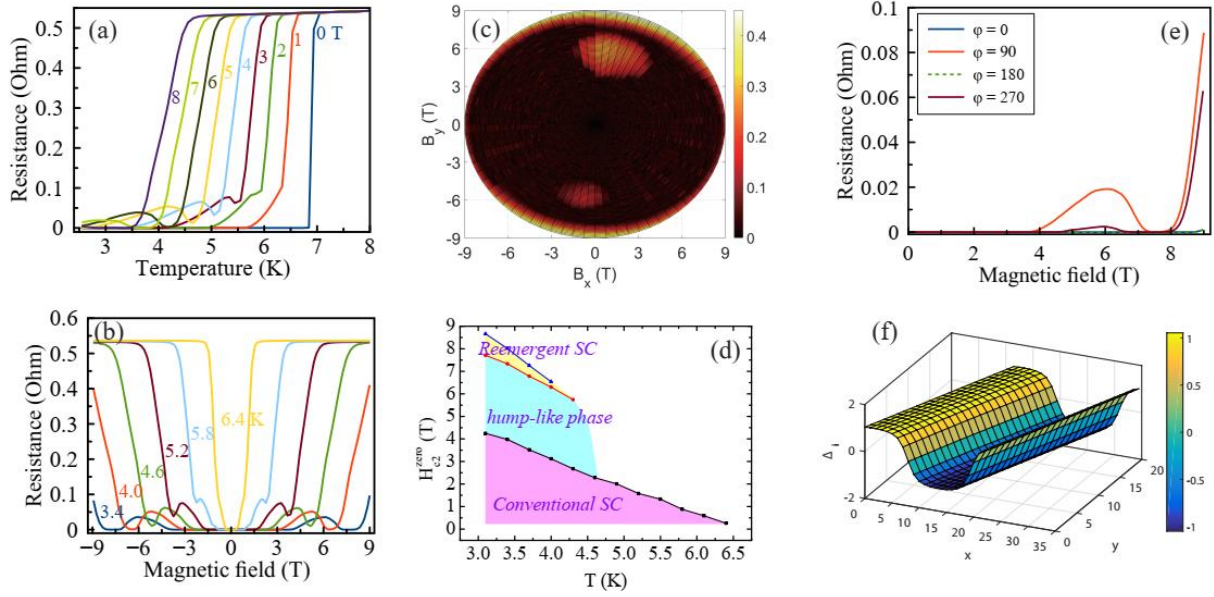

**Figure 3.** (a) The temperature dependent electric resistance of the NbSe$_2$/CrCl$_3$ heterostructure at various strengths of magnetic field that parallel to the sample plane but perpendicular to the electric current direction ($\theta$ = 90°, $\varphi$= 90°). (b) The corresponding MR curves at various temperatures. (c) The azimuthal angle $\varphi$ dependent MR curves of the NbSe$_2$/CrCl$_3$ heterostructure at $T$=3.4 K. Here the electric resistance magnitude is rescaled by cube root. (d) The experimental phase diagram of $H_{c2}^{zero} - T$ with the magnetic field parallel to the sample plane but perpendicular to the electric current direction ($\theta$ = 90°, $\varphi$= 90°). (e) The selected curves of (c) with azimuthal angle $\varphi$ of 0°, 90°, 180°, and 270°, respectively. (f) The theoretical calculation of the FFLO state driven by the strong spin-orbit coupling in the presence of in-plane magnetic field for the reemergent SC phase shown in (d).

## ASSOCIATED CONTENT

**Supporting Information**.

The following files are available free of charge.

FFLO state_si (PDF)

## AUTHOR INFORMATION

**Corresponding Author**

*D.J: jiangda@mail.sim.ac.cn

*W.L: w_li@fudan.edu.cn

**Author Contributions**

D.J. and W.L. conceived the project and designed the experiments. D.J. fabricated the samples and performed the experimental measurements. D.J., T.Y., W.L., G.M., and Z.A. analyzed the experimental data. Y.W. and W.L. did the theoretical model calculations. X.W. and W.L. performed the first-principles calculations. D.J. and W.L. wrote the paper. All authors have given approval to the final version of the manuscript.

‖ These authors contributed equally.

**Notes**

The authors declare no competing financial interest.

ACKNOWLEDGMENT

This work is supported by Superconducting ELectronics Facility (SELF) in SIMIT. We would like to acknowledge partial support by the Strategic Priority Research Program of Chinese Academy of Sciences (Grant No. XDB30000000), the National Natural Science Foundation of China (Grant Nos. 11927807), the Hundred Talents Program of Chinese Academy of Sciences, and the Natural Science Foundation of Shanghai of China (Grant Nos. 18JC1420402, 19ZR1402600, and 20DZ1100604). W.L. also acknowledges the start-up funding from Fudan University. All the authors are grateful for the crystal growth by G. Yu, the experimental assistance by H. Jin and L. Ma and graphic assistance by Q. Li.

REFERENCES

(1) Norman, M. R. The Challenge of Unconventional Superconductivity, *Science* **2011**, 332, 196-200.

(2) Séamus, J. C.; Lee, D. H. Concepts Relating Magnetic Interactions, Intertwined Electronic Orders, and Strongly Correlated Superconductivity, *Proc. Natl. Acad. Sci.* **2013**,110, 17623-17630.

(3) Stewart, G. R. Unconventional Superconductivity, *Advances in Physics* **2017**, 66, 75-196.

(4) Bednorz, J. G.; Müller, K. A. Possible High $T_c$ Superconductivity in the Ba-La-Cu-O System, *Zeitschrift für Physik B*, **1986**, 64, 189-193.

(5) Shen, Z.-X.; Dessau, D. S.; Wells, B. O.; King, D. M.; Spicer, W. E.; Arko, A. J.; Marshall, D.; Lombardo, L. W.; Kapitulnik, A.; Dickinson, P.; Doniach, S.; DiCarlo, J.; Loeser, T.; Park,

C. H. Anomalously Large Gap Anisotropy in the *a-b* Plane of $Bi_2Sr_2CaCu_2O_{8+\delta}$, *Phys. Rev. Lett.* **1993**, 70, 1553.

(6) Tsuei, C. C.; Kirtley, J. R.; Chi, C. C.; Yu-Jahnes, L. S.; Gupta, A.; Shaw, T.; Sun, J. Z.; Ketchen M. B. Pairing Symmetry and Flux Quantization in a Tricrystal Ring of Superconductin $YBa_2Cu_3O_{7-\delta}$, *Phys. Rev. Lett.* **1994**, 73, 593.

(7) Kamihara, Y.; Watanabe, T.; Hirano, M.; Hosono, H. Iron-based Layered Superconductor $La[O_{1-x}F_x]FeAs$ ($x$ = 0.05-0.12) with $T_c$ = 26 K, *J. Am. Chem. Soc.* **2008**, 130, 3296-3297.

(8) Mazin, I. I.; Singh, D. J.; Johannes, M. D.; Du, M. H. Unconventional Superconductivity with a Sign Reversal in the Order Parameter of $LaFeAsO_{1-x}F_x$, *Phys. Rev. Lett.* **2008**, 101, 057003.

(9) Ding, H.; Richard, P.; Nakayama, K.; Sugawara, K.; Arakane, T.; Sekiba, Y.; Takayama, A.; Souma, S.; Sato, T.; Takahashi, T.; Wang, Z.; Dai, X.; Fang, Z.; Chen, G. F.; Luo, J. L.; Wang, N. L. Observation of Fermi-surface-dependent Nodeless Superconducting Gaps in $Ba_{0.6}K_{0.4}Fe_2As_2$, *EuroPhysics Letters* **2008**, 83, 47001.

(10) Wang, F.; Zhai, H.; Ran, Y.; Vishwanath, A.; Lee, D.-H. Functional enormalization-group Study of the Pairing Symmetry and Pairing Mechanism of the FeAs-based High-temperature Superconductor, *Phys. Rev. Lett.* **2009**, 102, 047005.

(11) Fulde, P. and Ferrell, R. A., Superconductivity in a Strong Spin-exchange Field, *Phys. Rev.* **1964**, 135, A550-A563.

(12) Larkin, A. I. and Ovchinnikov, Y. N., Inhomogeneous State of Superconductors, *Sov. Phys. JETP* **1965**, 20, 762-770.

(13) Burkhardt, H.; Rainer, D. Fulde-Ferrell-Larkin-Ovchinnikov State in Layered Superconductors, *Ann. Physik* **1994**, 3, 181-194.

(14) Matsuda, Y.; Shimahara, H. Fulde-Ferrell-Larkin-Ovchinnikov State in Heavy Fermion Superconductors, *J. Phys. Soc. Jpn.* **2007**, 76, 051005.

(15) Lebed, A. G. The Physics of Organic Superconductors and Conductors (Springer, Berlin, **2008**).

(16) Zwierlein, M. W.; Schirotzek, A.; Schunck, C. H.; Ketterle, W. Fermionic Superfluidity with Imbalanced Spin Populations, *Science* **2006**, 311, 492-496.

(17) Liao, Y. A.; Rittner, A. S. C.; Paprotta, T.; Li, W.; Partridge, G. B.; Hulet, R. G.; Baur, S. K.; Mueller, E. J. Spin-imbalance in a One-dimensional Fermi Gas, *Nature* **2010**, 467, 567-569.

(18) Bianchi, A.; Movshovich, R.; Capan, C.; Pagliuso, P. G.; Sarrao, J. L. Possible Fulde-Ferrell-Larkin-Ovchinnikov Superconducting State in $CeCoIn_5$, *Phys. Rev. Lett.* **2003**, 91, 187004.

(19) Kenzelmann, M.; Strässle, Th.; Niedermayer, C.; Sigrist, M.; Padmanabhan, B.; Zolliker, M.; Bianchi, A. D.; Movshovich, R.; Bauer, E. D.; Sarrao, J. L.; Thompson, J. D. Coupled Superconducting and Magnetic Order in $CeCoIn_5$, *Science* **2008**, 321, 1652-1654.

(20) Kumagai, K.; Shishido, H.; Shibauchi, T.; Matsuda, Y. Evolution of Paramagnetic Quasiparticle Excitations Emerged in the High-field Superconducting Phase of CeCoIn$_5$, *Phys. Rev. Lett.* **2011**, 106, 137004.

(21) Wright, J. A.; Green, E.; Kuhns, P.; Reyes, A.; Brooks, J.; Schlueter, J.; Kato, R.; Yamamoto, H.; Kobayashi, M., Brown, S. E. Zeeman-driven phase transition within the superconducting state of κ-(BEDTTTF)$_2$Cu(NCS)$_2$, *Phys. Rev. Lett.* **2011**, 107, 087002.

(22) Mayaffre, H.; Krämer, S.; Horvatić, M.; Berthier, C.; Miyagawa, K.; Kanoda, K.; Mitrović, V. F. Evidence of Andreev Bound States as a Hallmark of the FFLO Phase in κ-(BEDTTTF)$_2$Cu(NCS)$_2$, *Nature Physics* **2014**, 10, 928-932.

(23) Koutroulakis, G.; Kühne, H.; Schlueter, J. A.; Wosnitza, J.; Brown, S. E. Microscopic Study of the Fulde-Ferrell-Larkin-Ovchinnikov State in an All-organic Superconductor, *Phys. Rev. Lett.* **2016**, 116, 067003.

(24) Lortz, R.; Wang, Y.; Demuer, A.; Böttger, P. H. M.; Bergk, B.; Zwicknagl, G.; Nakazawa, Y.; Wosnitza, J. Calorimetric Evidence for a Fulde-Ferrell-Larkin-Ovchinnikov Superconducting State in the Layered Organic Superconductor κ-(BEDTTTF)$_2$Cu(NCS)$_2$, *Phys. Rev. Lett.* **2007**, 99, 187002.

(25) Beyer, R.; Bergk, B.; Yasin, S.; Schlueter, J. A.; Wosnitza, J. Angle-dependent Evolution of the Fulde-Ferrell-Larkin-Ovchinnikov State in an Organic Superconductor, *Phys. Rev. Lett.* **2012**, 109, 027003.

(26) Agosta, C. C.; Fortune, N. A.; Hannahs, S. T.; Gu, S.; Liang, L.; Park, J.-H.; Schleuter, J. A. Calorimetric Measurements of Magnetic-field-induced Inhomogeneous Superconductivity above the Paramagnetic Limit, *Phys. Rev. Lett.* **2017**, 118, 267001.

(27) Uji, S.; Terashima, T.; Nishimura, M.; Takahide, Y.; Konoike, T.; Enomoto, K.; Cui, H.; Kobayashi, H.; Kobayashi, A.; Tanaka, H.; Tokumoto, M.; Choi, E. S.; Tokumoto, T.; Graf, D.; Brooks, J. S. Vortex Dynamics and the Fulde-Ferrell-Larkin-Ovchinnikov State in a Magnetic-field-induced Organic Superconductor, *Phys. Rev. Lett.* **2006**, 97, 157001.

(28) McGuire, M. A.; Clark, G.; Santosh K.C.; Chance, W. M.; Jellison, G. E. Jr.; Cooper, V. R.; Xu, X.; Sales, B. C. Magnetic Behavior and Spin-lattice Coupling Cleavable Van der Waals Layered $CrCl_3$ Crystals, *Phys. Rev. Mat.* **2017**, 1, 014001.

(29) Wang, X.; Song, Z.; Wen, W.; Liu, H.; Wu, J.; Dang, C.; Hossain, M.; Iqbal, M. A.; Xie, L. Potential 2D Materials with Phase Transitions: Structure, Synthesis, and Device Applications, *Adv. Mater.* **2018**, 31 (45), 1804682-1804695.

(30) Frindt, R. F. Superconductivity in Ultrathin $NbSe_2$ Layers, *Phys. Rev. Lett.* **1972**, 28, 299.

(31) Dvir, T.; Massee, F.; Attias, L.; Khodas, M.; Aprili, M.; Quay, C. H. L.; Steinberg, H. Spectroscopy of Bulk and Few-layer Superconducting $NbSe_2$ with Van der Waals Tunnel Junctions, *Nature Communications* **2018**, 9, 598.

(32) Ugeda, M. M.; Bradley, A. J.; Zhang, Y.; Onishi, S.; Chen, Y.; Ruan, W.; Ojeda-Aristizabal, C.; Ryu, H.; Edmonds, M. T.; Tsai, H.-Z.; Riss, A.; Mo, S.-K.; Lee, D.-H.; Zettl, A., Hussain, Z.; Shen, Z.-X.; Crommie, M. F. Characterization of Collective Ground States in Single-layer $NbSe_2$, *Nature Physics* **2016**, 12, 92-97.

(33) Buzdin, A. I. Proximity Effects in Superconductor-ferromagnet Heterostructures, *Rev. Mod. Phys.* **2005**, 77, 935.

(34) Michaeli, K.; Potter, A. C.; Lee, P. A. Superconductivity and Ferromagnetism in Oxide Interface Structures: Possibility of Finite Momentum Pairing, *Phys. Rev. Lett.* **2012**, 108, 117003.

(35) Hedo, M.; Kobayashi, Y.; Inada, Y.; Yamamoto, E.; Haga, Y.; Suzuki, J.; Metoki, N.; Ōnuki, Y.; Sugawara, H.; Sato, H.; Tenya, K.; Tayama, T.; Amitsuka, H.; Sakakibara, T. Peak Effect in CeRu$_2$: Role of Crystalline Defects, *J. Phys. Soc. Jpn.* **1998**, 67, 3561-3569.

(36) Tenya, T.; Yasunami, S.; Tayama, T.; Amitsuka, H.; Sakakibara, T.; Hedo, M.; Inada, Y.; Haga, Y.; Yamamoto, E.; Onuki, Y. Field-history-dependent Peak Effect in the Superconducting Mixed State of CeRu$_2$, *Physica B* **1999**, 259, 692-693.

(37) Jaccarino, V.; Peter, M. Ultra-high-field Superconductivity, *Phys. Rev. Lett.* **1962**, 9, 290.

(38) Uji, S.; Shinagawa, H.; Terashima, T.; Yakabe, T.; Terai, Y.; Tokumoto, M.; Kobayashi, A.; Tanaka, H.; Kobayashi, H. Magnetic-field-induced Superconductivity in a Two-dimensional Organic Conductor, *Nature* **2001**, 410, 908-910.

(39) Cao, Y.; Mishchenko, A.; Yu, G. L.; Khestanova, E.; Rooney, A. P.; Prestat, E.; Kretinin, A. V.; Blake, P.; Shalom, M. B.; Woods, C.; Chapman, J.; Balakrishnan, G.; Grigorieva, I. V.; Novoselov, K. S.; Piot, B. A.; Potemski, M.; Watanabe, K.; Taniguchi, T.; Haigh, S. J.; Geim, A. K.; Gorbachev, R. V. Quality Heterostructures from Two dimensional Crystals Unstable in Air by Their Assembly in Inert Atmosphere, *Nano Lett.* **2015**, 15, 4914-4921.

(40) Mineev, V. P.; Samokhin, K. V. Introduction to Unconventional Superconductivity, (Gordon and Breach Science Publishers, **1999**).

(41) Tinkham, M. Introduction to Superconductivity, 2nd edn (McGraw-Hill, New York, **1996**).

(42) Liu, J. M.; Zheliuk, O.; Leermakers, I.; Yuan, N. F. Q.; Zeitler, U.; Law, K. T.; Ye, J. T. Evidence for Two-dimensional Ising Superconductivity in Gated $MoS_2$, *Science* **2015**, 350, 1353-1357.

(43) Saito, Y.; Nakamura, Y.; Bahramy, M. S.; Kohama, Y.; Ye, J.; Kasahara, Y.; Nakagawa, Y.; Onga, M.; Tokunaga, M.; Nojima, T.; Yanase, Y.; Iwasa, Y. Superconductivity Orotected by Spinvalley Locking in Ion-gated $MoS_2$, *Nature Physics* **2016**, 12, 144-149.

(44) Xi, X.; Wang, Z.; Zhao, W.; Park, J.-H.; Law, K. T.; Berger, H.; Forró, L.; Shan, J.; Mak, K. F. Ising Pairing in Superconducting $NbSe_2$ Atomic Layers, *Nature Physics* **2016**, 12, 139-143.

(45) Guillamon, I.; Suderow, H.; Guinea, F.; Vieira, S. Intrinsic Atomic-scale Modulations of the Superconducting Gap of 2H-$NbSe_2$, *Phys. Rev. B* **2008**, 77, 134505.

(46) Tinkham, M. Effect of Fluxoid Quantumzation on Transitions of Superconducting Films, *Phys. Rev.* **1963**, 129, 2413.

(47) Zou, Y.-C.; Chen, Z.-G.; Zhang, E.; Xiu, F.; Matsumur, D.; Yang, L.; Hong, M.; Zou, J. Superconductivity and Magnetotransport of Single-crystalline $NbSe_2$ Nanoplates Grown by Chemical Vapour Deposition, *Nanoscale* **2017**, 9, 16591-16595.

(48) Mackenzie, A. P.; Maeno, Y. The Superconductivity of $Sr_2RuO_4$ and the Physics of Spin-triplet Pairing, *Rev. Mod. Phys.* **2003**, 75, 657.

(49) Almeida, D. E.; Fernandes, R. M.; Miranda, E. Induced Spin-triplet Pairing in the Coexistence State of Antiferromagnetism and Singlet Superconductivity: Collective Modes and Microscopic Properties, *Phys. Rev. B* **2017**, 96, 014514.

(50) Lévy, F.; Sheikin, I.; Grenier, B.; Huxley, A. D. Magnetic Field-induced Superconductivity in the Ferromagnet URhGe, *Science* **2005**, 309, 1343-1346.

(51) Ran, S.; Eckberg, C.; Ding, Q.-P.; Furukawa, Y.; Metz, T.; Saha, S. R.; Liu, I.-L.; Zic, M.; Kim, H.; Paglione, J.; Butch, N. P. Nearly Ferromagnetic Spin-triplet Superconductivity, *Science* **2019**, 365, 684-687.

(52) Knebel, G.; Knafo, W.; Pourret, A.; Niu, Q.; Vališka, M.; Braithwaite, D.; Lapertot, G.; Nardone, M.; Zitouni, A.; Mishra, S.; Sheikin, I.; Seyfarth, G.; Brison, J.-P.; Aoki, D.; Flouquet, J. Field-reentrant Superconductivity Close to a Metamagnetic Transition in the Heavy Fermion Superconductor $UTe_2$, *J. Phys. Soc. Jpn.* **2019**, 88, 063707.

(53) Wang, Y.-L.; Glatz, A.; Kimmel, G. J.; Aranson, I. S.; Thoutam, L. R.; Xiao, Z.-L.; Berdiyorov, G. R.; Peeters, F. M.; Crabtree, G. W.; Kwok, W.-K. Parallel Magnetic Field Suppresses Dissipation in Superconducting Nanostrips, *Proc. Natl. Acad. Sci.* **2017**, 114, E10274-E10280.

(54) Blatter, G.; Feigel'man, M. V.; Geshkenbein, V. B.; Larkin, A. I.; Vinokur, V. M. Vortices in High-temperature Superconductors, *Rev. Mod. Phys.* **1994**, 66, 1125.

(55) Zheliuk, O.; Lu, J. M.; Chen, Q. H.; El Yumin, A. A.; Golightly, S.; Ye, J. T. Josephson Coupled Ising Pairing Induced in Suspended MoS$_2$ Bilayers by Double-side Ionic Gating, *Nature Nanotechnology* **2019**, 14, 1123-1128.

(56) Mansky, P. A.; Chaikin, P. M.; Haddon, R. C. Evidence for Josephson Vortices in (BEDTTTF)$_2$Cu(NCS)$_2$, *Phys. Rev. B* **1994**, 50, 15929.

(57) Kirtley, J. R.; Moler, K. A.; Schlueter, J. A.; Williams, J. M. Inhomogeneous Interlayer Josephson Coupling in κ-(BEDT-TTF)$_2$Cu(NCS)$_2$, *J. Phys.: Condens. Matter* **1999**, 11, 2007-2016.

(58) Correa, A.; Mompeán, F.; Guillamón, I.; Herrera, E.; García-Hernández, M.; Yamamoto, T.; Kashiwagi, T.; Kadowaki, K.; Buzdin, A. I.; Suderow, H.; Munuera, C. Attractive Interaction between Superconducting Vortices in Tilted Magnetic Fields, *Commun. Phys.* **2019**, 2, 31.